# Electricity Market Theory Based on Continuous Time Commodity Model

Haoyong Chen[1], Lijia Han[2]

[1]Institute of Power Economics and Electricity Markets, South China University of Technology, Guangzhou 510641, China. Email: eehychen@scut.edu.cn

[2]Mathematical & Physical Science School, North China Electric Power University, Beijing 102206, China. Email: hljmath@ncepu.edu.cn

**Abstract**

The recent research report of U.S. Department of Energy prompts us to re-examine the pricing theories applied in electricity market design. The theory of spot pricing is the basis of electricity market design in many countries, but it has two major drawbacks: one is that it is still based on the traditional hourly scheduling/dispatch model, ignores the crucial time continuity in electric power production and consumption and does not treat the intertemporal constraints seriously; the second is that it assumes that the electricity products are homogeneous in the same dispatch period and cannot distinguish the base, intermediate and peak power with obviously different technical and economic characteristics. To overcome the shortcomings, this paper presents a continuous time commodity model of electricity, including spot pricing model and load duration model. The market optimization models under the two pricing mechanisms are established with the Riemann and Lebesgue integrals respectively and the functional optimization problem are solved by the Euler-Lagrange equation to obtain the market equilibria. The feasibility of pricing according to load duration is proved by strict mathematical derivation. Simulation results show that load duration pricing can correctly identify and value different attributes of generators, reduce the total electricity purchasing cost, and distribute profits among the power plants more equitably. The theory and methods proposed in this paper will provide new ideas and theoretical foundation for the development of electric power markets.

## 1. Introduction

In the letter of United States Secretary of Energy Rick Perry (Perry [2017]) to Federal Energy Regulatory Commission (FERC) on September 28, 2017, it is addressed that short-run markets may not provide adequate price signals to ensure long-term investments in appropriately configured capacity. Also, resource valuations tend not to incorporate superordinate network and/or social values such as enhancing resilience into resource or wires into investment decision making. The increased important of system resilience to overall grid reliability may require adjustments to market mechanisms that enable better valuation. This conclusion is quoted from Quadrennial Energy Review (QER [2017]). Furthermore, the current wholesale market price formation rules are also doubted. Rick Perry urges FERC to take immediate action to ensure that the reliability and resiliency attributes of generation with on-site fuel supplies are fully valued and develop new market rules that will achieve this urgent objective. In the U.S. Department of Energy's Staff Report to the Secretary (DOE [2017]), the problems with the current wholesale electricity markets and the relationship to reliability/resilience of power grids are investigated and several important findings are reported. It is suggested that FERC should expedite its efforts with states, RTO/ISOs, and other stakeholders to improve energy price formation in centrally-organized wholesale electricity markets. Energy price formation reform is supported after several years of fact finding and technical conferences. DOE staff identified several research topics including market structure and pricing mechanism for enabling equitable, value-based remuneration for desired grid attributes.

The definition of electricity commodity model, cost structure and price formation mechanism are the most basic problems regardless of market structure. As the physical indifference (electricity produced by the power plants cannot be separated once injected into the power grid), and the existence of the complex physical network (power system), electricity becomes the world's most complex commodity. So the definition and pricing theory of electricity commodity are not so obvious. Spot Pricing of Electricity published in 1988 by Prof. F. C. Schweppe of the Massachusetts Institute of Technology is the classic literature of electricity pricing theory and became the theoretical basis of spot electricity market design in different countries (Schweppe [1988]). Electricity price theory research should be divided into two parts, namely, electricity cost analysis (that is, what is a reasonable price) and electricity price formation mechanism in electricity markets. In the ideal electricity market, the clearing price should be equal to the marginal cost of power plants and marginal utility of power users.

The theory of spot pricing is delicate in mathematics, but it does not conform to the physical characteristics of electricity production and consumption. In reality, the wholesale markets based on spot pricing theory has more or less problems. Besides those reported in DOE (2017), spot price often changes dramatically, brings great financial risk to market participants, financial instruments and other hedging measures are indispensable, which are often accompanied with unfair arbitrage; most power users are incapable (or unintentional) to respond to the rapidly changing real-time electricity price and rely on retailers to convert the real-time electricity price of the wholesale market into retailing packages with simple price structure, and thus the theoretical objective of enhancing demand-supply interaction through responsive spot pricing is not achieved; spot prices cannot completely cover the investment cost and result in insufficient investment capacity, etc.

**2. Principle of Spot Pricing**

The hourly spot price is determined by the supply/demand conditions that exist at that hour (Schweppe [1988]).

In particular, it depends on that hour's:
- demand (in total and by location);
- generation availability and costs (including purchases from other utilities);
- transmission/distribution network availability and losses.

The hourly spot price (without revenue reconciliation) is given by the marginal cost:

$$\rho_k(t) = \frac{\partial}{\partial d_k(t)}[\text{Total cost of providing energy to all customers now and through the future}]$$

where $\rho_k(t)$ is hourly spot price for $k$th customer during hour $t$ ($/kWh); $d_k(t)$ is demand of $k$th customer hour $t$ (kWh), subject to constraints such as:
- **energy balance**: total generation equals total demand plus losses;

- **generation limits**: total demand during hour t cannot exceed the capacity of all the power plants available at hour t;
- **Kirchoff's laws**: energy flows and losses on a network are specified by physical laws;
- **line flow limits**: energy flows over a particular line cannot exceed specified limits without causing system operating problems.

Spot price is formed on basis of the principle of social welfare maximization in classical microeconomics. In its original theory, short-term and long-term, operation and planning are taken into account (Bohn [1982]). In real-world electricity markets, spot price is often calculated by Security Constrained Unit Commitment (SCUC), Security Constrained Economic Dispatch (SCED) and other short-term operation optimization model. However, the spot price calculation still adopts the hourly power balance model of the traditional economic dispatch, which divides the whole trading interval into a series of cycles, and further divides each cycle into several periods, and then the energy balance model is calculated period by period. Since the power (system demand) for each period is assumed to be constant, the energy balance model is equivalent to the power balance model. As shown in Fig. 1, in the calculation, the electricity commodity model is in fact defined as follows:

The area under the entire load curve is divided into several "slips" by trading period, and then each slip is divided into several "segments". In each period, each winning bidder (unit) takes a "segment" (that is, a commodity), and in marginal price clearing mechanism, the settlement price for all commodity in the same trading period is the same (that is, the highest price of the winning unit).

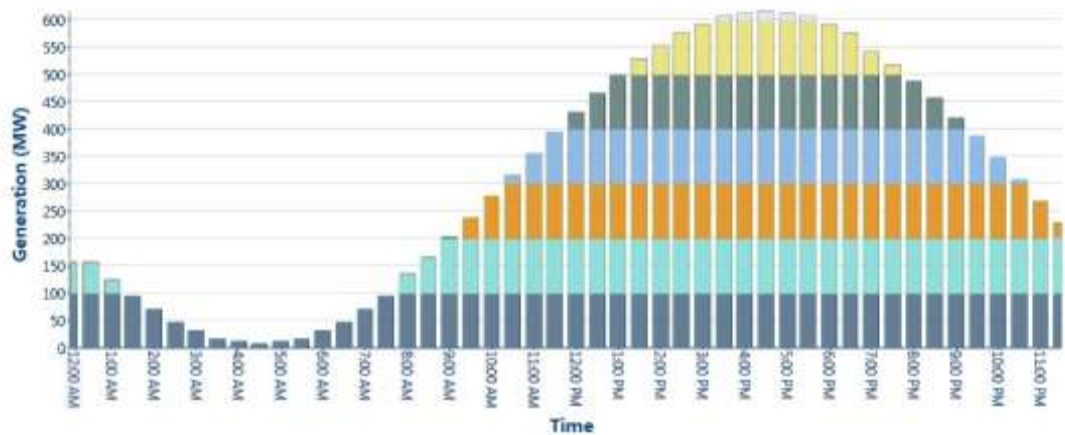

**Fig. 1 Electricity commodity model in spot pricing theory**

### 3. The continuous time electricity commodity model

Electricity markets based on spot pricing have the following shortcomings.

1) Spot price calculation is based on traditional hourly dispatch (or power balance) model, and therefore overlooks the crucial time continuity in electric power production and consumption, which is of special importance with large-scale integration of wind, solar and other renewable energies into power systems and the sharp rising requirements of flexibility; the inter-temporal constraints are not considered seriously (although they are mentioned in Schweppe [1988] but not investigated thoroughly).

2) Formation of spot price has a implied hypothesis that the electricity commodity in the same trading period are homogeneous, and then the technical characteristics and cost structure of the base load, intermediate load and peak load cannot be distinguished because the time dynamic characteristics of the different types of generator output are not considered.

3) Although the original spot pricing model covers resource optimization from operation to planning over a long horizon (Bohn [1982]), such a large-scale optimization problem cannot be applied in practice. The actual market often uses Security Constrained Unit Commitment (SCUC) or Security Constrained Economic Dispatch (SCED) models to calculate the clearing prices, which cannot give price signals reflecting long-term capacity investment, and then cannot guarantee the adequacy of generation capacity.

4) The production and consumption of electricity has the distinguishing feature of time continuity, and the commodity for sale or purchase is an "energy block" with certain duration for both the generator and the consumer. Energy (in MWh) is the main concern for power producers and users, and power (in MW) balance is mainly used as the physical constraint of power system operation (mainly in power flow equation). In the spot pricing theory, because time is not included in the commodity model, power balance and energy balance are equivalent, the physical balance constraints are directly used as the market equilibrium conditions, and this is not appropriate in economics.

5) Energy type electricity commodity has characteristics different from power type electricity commodity. Energy type commodity is more similar to ordinary commodity. In long term transactions energy type commodity actually can be stored, mainly in form of coal (or other fuel) and reservoir storage, which is in contrast to power type commodity which cannot be stored.

6) Electricity as a kind of the basic social product and means of production, the most critical consideration is to ensure sustained and stable supply. Society places value on attributes of electricity provision beyond those compensated by the current design of the wholesale market, such as jobs, community economic development, low emissions, local tax payments, etc. The highest market efficiency (especially the short-term spot market efficiency) is not so important.

In order to emphasize the role of time in electricity commodity, this paper redefines the continuous time commodity model of electricity $(P,t)(t_1 \leq t \leq t_2)$ by (power, time) pair, and the area under the power curve is energy, as shown in Fig. 2. When $t_2 = t_1 + 1$ and $P = \mathrm{const}(t_1 \leq t \leq t_2)$, it is degraded into hourly electricity commodity model in spot pricing theory. Since power can be viewed as a function of time, the continuous time electricity commodity model can be written as $(P(t),t)(t_1 \leq t \leq t_2)$, and the numerical power value in the spot price definition becomes a function defined in the time interval $(t_1,t_2)$. Mathematical theories, such as functional analysis, variational method, etc. are needed for analysis. After the introduction of the continuous time electricity commodity model, the problem of social welfare maximization changes from the multi-stage static optimization to the continuous time functional optimization. The solution method is also changed from Lagrange method to solution of Euler-Lagrange equation.

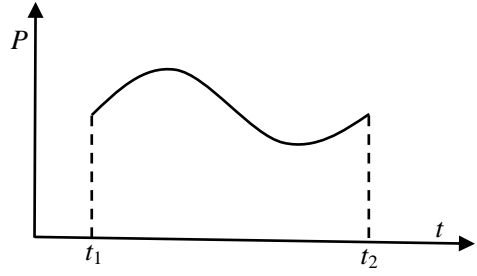

**Fig.2 Continuous time electricity commodity model**

## 4. Modelling of electricity market based on continuous time commodity

The microeconomic model of social welfare maximization in Caramanis (1982) is still used in this paper to define the electricity market for reference. Some variable notations and expressions in Caramanis (1982) are still used. Because the purpose of this paper is mainly to explain the basic principles, the market model in Caramanis (1982) is simplified.

**Market participant cost/benefit**

Supposing $B_j (j \in \phi)$ to be the variable cost (benefit) of market participant $j$ in a market cycle, the cost (benefit) function can be written as:

$$B_j = B_j\left(P_j(t)\right) \ (j \in \phi, 0 \leq t \leq T)$$

where $\phi$ is the set of all participants in the market; $B_j$ is a functional of power curve $P_j(t)$ for the market participant $j$, $B_j \leq 0$ denotes the generation cost when $j$ is a power plant and $B_j \geq 0$ denotes the consumption utility when $j$ is a power user; T is the duration of the market cycle.

**Social welfare and market objective**

The objective function is the standard social welfare maximization of the market in microeconomics, namely, to maximize

**social welfare = the value of electricity consumption - the cost of electricity production**

The objective function can be written as

$$\max \ W(P_1, P_2 \cdots P_n) = \sum_{j \in \phi} B_j(P_j(t))dt \tag{1}$$

**Constraints**

The key constraint is the power balance constraint, namely,

**0 = generation – losses - consumption**

which can be written as

$$0 = e(t) = \sum_j P_j(t) - L(t) \ \left(j \in \phi, 0 \leq t \leq T, L(t) > 0\right) \tag{2}$$

Note that time $t$ in (2) is a continuous variable and no longer discretized as in traditional economic dispatch. $P_j(t) \geq 0$ when $j$ is a power plant, and otherwise $P_j(t) \leq 0$ when $j$ is a power user. $L(t)$ is the loss of active power.

**Participant behavior**

With the continuous time electricity commodity model, the profit maximization model of market participants will take the form of integral (or functional of the power curve). The profit maximization model under spot pricing corresponds to the Riemann (Riemann) integral in mathematics, and the profit maximization model under load duration (called "measure" in mathematics) pricing is introduced in this paper, which corresponds to the Lebesgue integral. Note that the basic idea of load duration pricing has been described in Elmaghraby (1999). The basics of Riemann and Lebesgue integral are introduced in the appendix.

1) The participant profit maximization model under spot pricing (Riemann integral form)

Assuming that participant $j$ obtains income (for a power plant) or pays purchasing cost (for a power user) at spot price $\tilde{\pi}_j(t)(0 \leq t \leq T)$ changing over time, the participant's objective is to maximize the net profit under production (consumption) capacity constraints, namely

$$\max \quad N_j^R(P_j(t)) = \int_{t=0}^{T} [B_j(P_j(t)) + \tilde{\pi}_j(t) P_j(t)] dt \qquad (3)$$
$$s.t. \quad P_j^{\min}(t) \leq P_j(t) \leq P_j^{\max}(t)$$

For a particular participant, $P_j^{\min}(t), P_j^{\max}(t)$ are often constants (not changing over time).

2) The participant profit maximization model under load duration pricing (Lebesgue integral form)

Assuming that participant $j$ obtains income (for a power plant) or pays purchasing cost (for a power user) at the price $\tilde{\pi}_j(y)\left(P_j^{\min} \leq y \leq P_j^{\max}\right)$ determined by load duration (the load curve is assumed to be monotonically increasing in this paper, so the price function can also be written as a function of the load), the participant's objective is to maximize the net profit under production (consumption) capacity constraints, namely

$$\max \quad N_j^L(m_j(y)) = \int_{P_j^{\min}}^{P_j^{\max}} [B_j(y) \pm \tilde{\pi}_j(m_j(y)) m_j(y)] dy \qquad (4)$$
$$s.t. \quad P_j^{\min}(t) \leq P_j(t) \leq P_j^{\max}(t)$$

where $m_j(y) = m\{t : P_j(t) > y\}$ is the measure of load function at value $y$; the sign before $\tilde{\pi}_j(m_j(y))$ takes "+" sign when $j$ is a power plant and takes "-" sign when $j$ is a power user.

**Commodity model**

By the two different pricing methods, the models of electricity commodity are also different. Under spot pricing, the power curve $P_j(t)(j \in \phi, t_1 \leq t \leq t_2)$ of participant $j$ within a time horizon $t(t_1 \leq t \leq t_2)$ is regarded as one item of commodity, and the total price is $\int_{\tau=t_1}^{t_2} \tilde{\pi}_j(\tau) P_j(\tau) d\tau$. Then we can further define the per unit electric energy price of the commodity:

$$\tilde{\pi}_j = \frac{\int_{\tau=t_1}^{t_2} \tilde{\pi}_j(\tau) P_j(\tau) d\tau}{\int_{\tau=t_1}^{t_2} P_j(\tau) d\tau} \tag{5}$$

Under load duration pricing, the "energy block" of the participant $j$ within a certain power range $P_j(t)(P_1 \leq t \leq P_2)$ is considered to be one item of commodity, i.e.

$$\bar{P}_j(t) = \begin{cases} P_j(t) : P_1 \leq P_j(t) \leq P_2 \\ P_2 : P_j(t) > P_2 \end{cases} \left(j \in \phi, t \in \{t : P_j(t) \geq P_1\}\right),$$

Its total price is $\int_{P_1}^{P_2} \tilde{\pi}_j(y) m_j(y) dy$, and the per unit electric energy price of the commodity is:

$$\tilde{\pi}_j = \frac{\int_{P_1}^{P_2} \tilde{\pi}_j(y) m_j(y) dy}{\int_{P_1}^{P_2} m_j(y) dy} \tag{6}$$

**Market mechanism**

The objective of the overall market optimization is to maximize the social welfare in (1) subject to the constraints in (2). It is obvious that the problem is a variational problem, and according to its optimality condition, the dual variable (shadow price) $\lambda(t)$ can be obtained.

In the market mechanism based on spot pricing, $\lambda(t)$ is directly used as the market price $\tilde{\pi}_j(t)$ for each participant $j$, and that is, $\tilde{\pi}_j(t) = \lambda(t)(j \in \phi)$. $\lambda(t)$ is the market equilibrium price, and at this price the individual's optimal of (3) is also the solution of the overall market optimization problem (1).

In the market mechanism based on load duration, it is necessary to find the market price $\tilde{\pi}_j(m_j(y))$ for participant $j$, which satisfies $\tilde{\pi}_j(m_j(y)) = \pi(m(y))(j \in \phi)$. $\pi(m(y))$ is the market equilibrium price, and at this price the individual's optimal of (4) is also the solution of the overall market optimization problem (1). A specific market model and case study is used to expound the details of the new market modelling theory.

5. **Solution of electricity market model based on continuous time commodity**

For convenience of solution and analysis, this paper considers the unilateral competition model of generation side electricity market. Assuming that the cost function of power plants takes the form of quadratic function and ignoring network loss, the problem of social welfare maximization in (1) and (2) becomes generation cost minimization, namely,

$$\min\ C(P_1, P_2 \cdots P_n) = \sum_{j=1}^{n} \int_0^T C_j(P_j(t))dt$$
$$= \sum_{j=1}^{n} \int_0^T \left(a_j P_j(t)^2 + b_j P_j(t) + c_j\right)dt \qquad (7)$$
$$s.t.\ \sum_{j=1}^{n} P_j(t) = P_d(t)\ (0 \le t \le T)$$

where $P_d(t)$ is the system load at time $t$.

First, the variational problem with one constraint (7) is solved, and its Euler-Lagrange equation is

$$\begin{cases} C_1'(P_1(t)) - \lambda(t) = 0 \\ C_2'(P_2(t)) - \lambda(t) = 0 \\ \quad \vdots \\ C_n'(P_n(t)) - \lambda(t) = 0 \\ \sum_{j=1}^{n} P_j(t) - P_d(t) = 0 \end{cases} \qquad (8)$$

From the above equation j we can express $P_j(t)$ $(j=1,\cdots,n)$ by $\lambda(t)$. Then by substituting these expression of $P_1(t)$ and $P_n(t)$ into the power balance equation in (7), $\lambda(t)$ can be solved out, and then all $P_1(t),\ldots,P_n(t)$ can be got.

Then, the variational problem (3) in the sense of Riemann integral is solved as follows. By ignoring the capacity constraints, (3) is regarded as an unconstrained variational problem, and the Euler-Lagrange equation is

$$C_j'\left(P_j(t)\right) - \tilde{\pi}_j(t) = 0 \qquad (9)$$

By comparing (8) and (9), we can get

$$\tilde{\pi}_j(t) = \lambda(t) \qquad (10)$$

Namely $\lambda(t)$ is the market equilibrium price.

When the generation cost function takes the form of quadratic function, we have

$$\lambda(t) = C_j'\left(P_j(t)\right) = 2a_j P_j(t) + b_j = \frac{P_d(t)}{\sum_{j=1}^{n} \frac{1}{2a_j}} + \frac{\sum_{j=1}^{n} \frac{b_j}{2a_j}}{\sum_{j=1}^{n} \frac{1}{2a_j}} \qquad (11)$$

which shows the consistence of spot price with load curve.

Finally, the variational problem (4) in the Lebesgue integral sense is solved. Under the assumption of monotonically increasing load curve and unilateral competition of generators, (4) can be written as follows:

$$\min\ N_j^L(P_j) = \int_{P_{\min}}^{P_{\max}} [(T - P_j^{-1}(y))C_j'(y) - \tilde{\pi}_j(T - P_j^{-1}(y))m_j(y)]dy \qquad (12)$$

where $m_j(y) = m\{t : P_j(t) > y\}$ is the measure of $m_j(y)$ at value $y$.

$$P_{\max} = \max_{t \in [0,T]}\{P(t)\},\ P_{\min} = \min_{t \in [0,T]}\{P(t)\}$$

where

$$\int_{P_{\min}}^{P_{\max}} [T - P_j^{-1}(y)] C_j'(y) dy + C_j(P_{\min}) T \overset{h=C_j(y)}{=} \int_{C_j(P_{\min})}^{C_j(P_{\max})} [T - P_j^{-1} C_j^{-1}(h)] dh + C_j(P_{\min}) T$$
$$= \int_{t=0}^{T} C_j(P_j(t)) dt$$

is the generation cost.

When the cost function takes the quadratic form, it can be seen from (11) that $P_j(t)$ is a strictly monotonic increasing function, and there is $m_j(P_j(t)) = T - t$. We can use variable substitution $P_j(t) = y$, then (12) turns into the following variational problem:

$$\max \ N_j^L(P_j) = \int_0^T [(T-t) C_j'(P_j(t)) - \tilde{\pi}_j(t) m_j(P_j(t))] P'_j(t) dt$$
$$= \int_0^T [(T-t)(2a_j P_j(t) + b_j) - \tilde{\pi}_j(t)(T-t)] P'_j(t) dt \quad (13)$$

The Euler-Lagrange is

$$2a_j P'_j(t)(T-t) + \frac{d}{dt}[(T-t)(2a_j P_j(t) + b_j) - \tilde{\pi}_j(t)(T-t)] = 0 \quad (14)$$

By solving the first order linear ordinary differential equation (14), we can get the price function $\tilde{\pi}_j(t)$.

On the other hand, from (11) there is

$$P_j(t) = \frac{1}{2a_j} \left( \frac{P_d(t)}{\sum_{j=1}^n \frac{1}{2a_j}} + \frac{\sum_{j=1}^n \frac{b_j}{2a_j}}{\sum_{j=1}^n \frac{1}{2a_j}} - b_j \right)$$

It is easy to find that the coefficient and nonlinear term of (14)

$$2a_j P'_j(t) = \frac{P_d'(t)}{\sum_{j=1}^n \frac{1}{2a_j}}, \quad 2a_j P_j(t) + b_j = \lambda(t)$$

are irrelevant to $j$, and then the solution of (14) is relevant to $j$, which means that the prices for all power plants are the same at the optimum, that is,

$$\tilde{\pi}_j(t) = \pi(t) \quad (15)$$

Namely $\pi(t)$ is the market equilibrium price.

6. **Distribution of commodity values under two different pricing mechanism**

"*Same quality same price*" is the basic principle for commodity pricing. Under spot pricing mechanism, the underlying assumption is that all electricity commodities at the same trading period are homogeneous and therefore have the same marginal price (or value); and the electricity commodities at different trading periods are heterogeneous and have different prices (values), as shown in Fig. 3. For the same power plant (i.e., the cross bar consisting of multiple small segments of the same color in Fig.

3), the value of the electricity commodities produced over time is different. This does not conform to the actual operation of power systems. At the same period (vertical bar consisted of segments with different colors), all electricity commodities are homogeneous. Then the obviously difference in technical characteristics and cost composition among the base load, intermediate load and peak load cannot be distinguished. This is one of the deficiencies of spot pricing, which cannot correctly value different attributes provided by different generators.

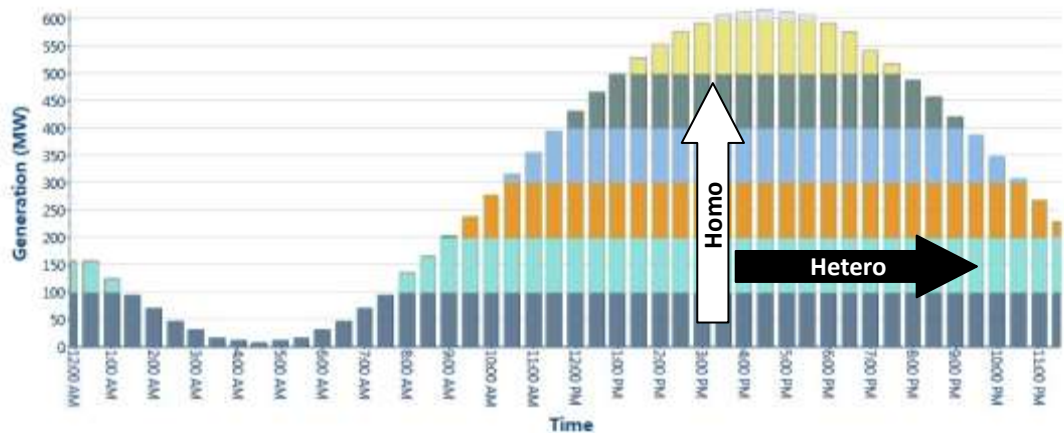

**Fig.3 Distribution of commodity values under spot pricing**

Under load duration pricing mechanism presented in this paper, all electricity commodities with the same system load duration (namely on a horizontal line representing a specific load level in Fig. 4) are homogeneous and therefore have the same marginal price (or value). As system load duration changes, the commodity price changes accordingly. The shorter the load duration, the higher the price (its economic meaning is that the peak load price is higher than the base load price). As shown in Fig. 4, for the same power plant (i.e., the cross bar with the same color), the value of the electricity commodities produced over time is the same. On the other hand, at the same period (vertical bar consisted of segments with different colors), all electricity commodities are heterogeneous, and the attributes and prices of power produced by the base load, intermediate load and peak load units are different.

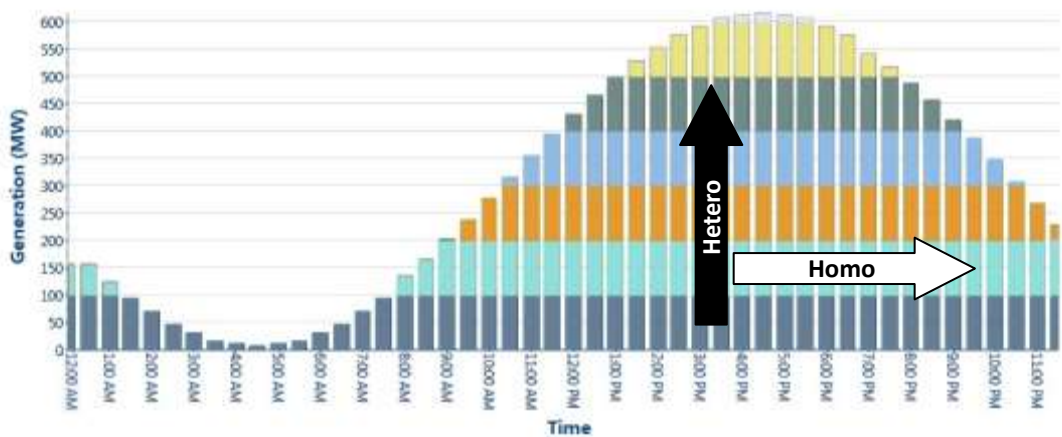

**Fig.4 Distribution of commodity values under load duration pricing**

7. **Case study**

Assuming the load curve is monotonically increasing and as follows
$$P_d(t) = 350 \times (1+2t) \quad (t \in [0, 1])$$

Assuming further that there are 3 power plants in the market, and the generation cost is different by 2 times successively, representing the low, medium and high cost power plant respectively. The cost function is as follows (where $a = 0.001, b = 0.07, c = 0.2$)

$$C_1(P_1(t)) = \frac{a}{2}P_1(t)^2 + bP_1(t) + c$$

$$C_2(P_2(t)) = aP_2(t)^2 + 2bP_2(t) + 2c$$

$$C_3(P_3(t)) = 2aP_3(t)^2 + 4bP_3(t) + 4c$$

From (8), the spot price and generation output of each power plant are solved as follows,

$$\lambda(t) = C_1'(P_1(t)) = aP_1(t) + b, \quad P_1(t) = \frac{4}{7}P_d(t) + \frac{5b}{7a} = 250 + 400t,$$

$$\lambda(t) = C_2'(P_2(t)) = 2aP_2(t) + 2b, \quad P_2(t) = \frac{2}{7}P_d(t) - \frac{b}{7a} = 90 + 200t,$$

$$\lambda(t) = C_3'(P_3(t)) = 4aP_3(t) + 4b, \quad P_3(t) = \frac{1}{7}P_d(t) - \frac{4b}{7a} = 10 + 100t,$$

The dispatch schemes of the 3 power plants and system load are shown in Fig. 5. The spot price is shown in Fig. 6 (a). Obviously it is linear as the system load.

The generation costs of 3 power plants can be obtained:

$$C_1 = \int_0^1 \left(\frac{a}{2}P_1(t)^2 + bP_j(t) + c\right)dt$$
$$= \int_0^1 \left(\frac{a}{2}(250+400t)^2 + b(250+400t) + c\right)dt$$
$$= 107.92 + 450b + c = 139.63$$

$$C_2 = \int_0^1 \left(aP_2(t)^2 + 2bP_2(t) + 2c\right)dt$$
$$= \int_0^1 \left(a(90+200t)^2 + 2b(90+200t) + 2c\right)dt$$
$$= 39.43 + 380b + 2c = 66.46$$

$$C_3 = \int_0^1 \left(2aP_3(t)^2 + 4bP_3(t) + 4c\right)dt$$
$$= \int_0^1 \left(2a(10+100t)^2 + 4b(10+100t) + 4c\right)dt$$
$$= 8.86 + 240b + 4c = 26.44$$

1) Analysis of revenue and profit under spot pricing

The revenues of 3 power plants can be calculated as

$$N_1^R(P_1) = \int_{t=0}^1 \lambda(t)P_1(t)dt = \int_{t=0}^1 [0.32+0.4t](250+400t)dt = 247.32$$

$$N_2^R(P_2) = \int_{t=0}^1 \lambda(t)P_2(t)dt = \int_{t=0}^1 [0.32+0.4t](90+200t)dt = 105.52$$

$$N_3^R(P_3) = \int_{t=0}^{1} \lambda(t) P_3(t) dt = \int_{t=0}^{1} [0.32 + 0.4t](10 + 100t) dt = 34.48$$

The profits of 3 power plants are respectively 107.69, 39.06, 8.04, which are shown in Fig. 7(a). The profit rates ( profit / cost×100% ) are respectively 77%, 59%, 30%. The total generation cost and total profit of 3 power plants is 232.53 and 154.79 respectively. The total electricity purchasing cost and profit rate of the market are 387.32 and 67% ( total profit / total cost×100% ) respectively.

2) Analysis of market clearing price, revenue and profit under load duration pricing

Because $m_j(P_j(t)) = 1-t$, $a_j P'_j(t) = 0.2$ $(j=1,2,3)$, by substituting the example data into the corresponding Euler-Lagrange equation (14), we can get:

$$2a_j(1-t)P'_j(t) + \frac{d}{dt}[(1-t)(2a_j P_j(t) + b_j) - \tilde{\pi}_j(t)(1-t)] = 0$$

It can be arranged to ordinary differential equation:

$$\tilde{\pi}'_j(t)(1-t) - \tilde{\pi}_j(t) + 1.2t - 0.48 = 0 \quad (j=1,2,3)$$

The solution is

$$\tilde{\pi}_j(t) = Ce^{\int \frac{1}{1-t}} + e^{\int \frac{1}{1-t}} \int (1.2 - \frac{0.72}{1-t}) e^{-\int \frac{1}{1-t}} dt$$
$$= C(1-t)^{-1} + (1-t)^{-1} \int [1.2(1-t) - 0.72] dt \quad (t \in [0,1), \; j=1,2,3)$$
$$= C(1-t)^{-1} + (1-t)^{-1}(0.48t - 0.6t^2)$$

Let $\tilde{\pi}_j(0) = \lambda(0)$, then $C = 0.32$, and we have

$$\pi(t) = \tilde{\pi}_j(t) = 0.32(1-t)^{-1} + (1-t)^{-1}(0.48t - 0.6t^2) \quad (t \in [0,1), \; j=1,2,3)$$

Because the measurement $m = 1-t$, then

$$\pi_j(m) = \tilde{\pi}_j(1-t) = 0.32m^{-1} + m^{-1}(0.48(1-m) - 0.6(1-m)^2) \quad (m \in (0,1], \; j=1,2,3)$$

By substituting the example data into (12), we can get the corresponding measurement functions:

$$m_1(y) = \frac{650-y}{400}, \quad y \in [250, 650]$$
$$m_2(y) = \frac{290-y}{200}, \quad y \in [90, 290]$$
$$m_3(y) = \frac{110-y}{100}, \quad y \in [10, 110]$$

Thus the revenues of 3 power plants can be calculated as:

$$N_1^L(P_1) = \int_{250}^{650} \pi_1(m_1(y)) m_1(y) dy + 0.32 \times 250$$
$$= \int_{250}^{650} [0.32 + (0.48(1-m_1(y)) - 0.6(1-m_1(y))^2)] dy + 0.32 \times 250$$
$$= \int_{250}^{650} [0.2 + 0.72 m_1(y) - 0.6 m_1(y)^2] dy + 80$$
$$= 224$$

$$N_2^L(P_2) = \int_{90}^{290} \pi_2(m_2(y))m_2(y)dy + 0.32 \times 90$$
$$= \int_{90}^{290}[0.4 + 0.72m_2(y) - 0.8m_2(y)^2\ ]\,dy + 98.8$$
$$= 100.84$$

$$N_3^L(P_3) = \int_{10}^{110} \pi_3(m_3(y))m_3(y)dy + 0.32 \times 10$$
$$= \int_{10}^{110}[0.4 + 0.72m_3(y) - 0.8m_3(y)^2\ ]\,dy + 3.2$$
$$= 39.16$$

Here

$$\int_{250}^{650} m_1(y)dy = \int_{250}^{650}\frac{650-y}{400}dy = 200$$

$$\int_{250}^{650} m_1(y)^2 dy = 133.33$$

$$\int_{90}^{290} m_2(y)\,dy = \int_{90}^{290}\frac{290-y}{200}\,dy = 100$$

$$\int_{90}^{290} m_2(y)^2\,dy = 66.67$$

$$\int_{10}^{110} m_3(y)\,dy = \int_{10}^{110}(\frac{110-y}{100})\,dy = 50$$

$$\int_{10}^{110} m_3(y)^2\,dy = 33.33$$

The load duration price is shown in Fig. 6(b), which is nonlinear since it is related to the duration of system load. The profits of 3 power plants are respectively 84.37, 34.38, 12.72, which are shown in Fig. 7(b). The profit rates are respectively 60%, 52%, 48%. The total generation cost and total profit of 3 power plants is 232.53 and 131.47 respectively. The total electricity purchasing cost and profit rate of the market are 364 and 57% respectively.

From Fig. 7 we can see that because spot pricing mechanism cannot distinguish base load, intermediate load and peak load, the profits of 3 power plants differ greatly. Power plant 1 is allocated too much profit. Load duration pricing can reduce the total electricity purchasing cost, and the profit allocation among the power plants is more equitable. Compared with spot pricing mechanism, the profit of power plant 1 is reduced and that of power plant 3 is increased.

As the two pricing mechanisms have their own pros and cons, the future research will consider combination of the two pricing mechanisms.

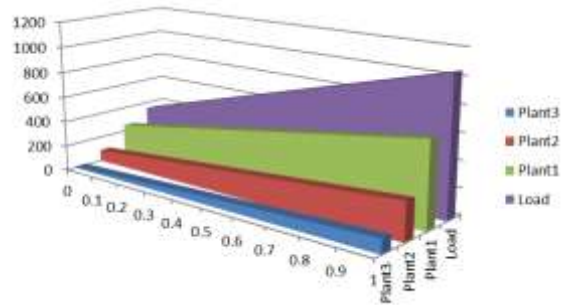

**Fig.5 Distribution of commodity values under load duration pricing**

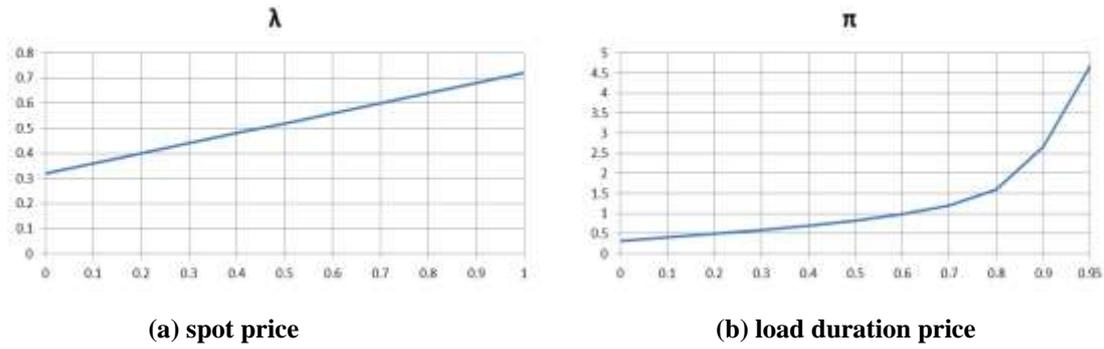

| (a) spot price | (b) load duration price |

Fig.6 Prices under two different pricing mechanism

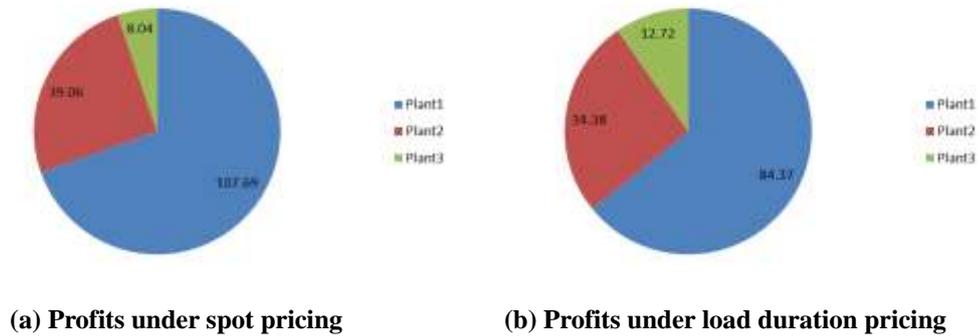

| (a) Profits under spot pricing | (b) Profits under load duration pricing |

Fig.7 Profits under two different pricing mechanism

## 8. Conclusion

In the letter of Secretary of Energy Rick Perry to FERC, it is highlighted that distorted price signals in the Commission-approved organized markets have resulted in under-valuation of grid reliability and resiliency benefits provided by traditional baseload resources. The Commission has recognized that there are deficiencies in the way the regulated wholesale power markets price power and that these deficiencies are undermining reliability and resiliency. So it is the Commission's immediate responsibility to take action to ensure that the reliability and resiliency attributes of generation with on-site fuel supplies are fully valued and in particular to exercise its authority to develop new market rules that will achieve this urgent objective. In particular, the value of on-site fuel storage capability must be accounted for.

To achieve these goals, the basic theories of electricity commodity and market mechanism are urgently needed for further study. Beginning with the analysis of wholesale electricity markets based on spot pricing theory, this paper presents the continuous time electricity commodity models, including commodity models under spot pricing and load duration pricing, based on which the market models of social welfare maximization are established. The market optimization models under spot pricing and load duration pricing correspond to Riemann integral and Lebesgue integral in mathematics respectively. The market equilibrium solution is obtained by solving the respective Euler-Lagrange equation. In particular, the feasibility of load duration pricing is proved by strict mathematical derivation. Finally, an example is given to verify the correctness of the theories and methods proposed. This paper is expected to provide novel ideas and theoretical basis for electricity market design.

## 9. Appendix

The basics of Riemann integral (Zaidman [1999]) and Lebesgue integral (Bear [1988]) are introduced as follows.

**1) Riemann integral**

Giving a bounded function $f(x)$ on $[a,b]$ and a dissection $\sigma$ of $[a,b]$, $a = x_0 < x_1 < x_2 < \cdots < x_{n-1} < x_n = b$.

Then the interval $[a,b]$ is devided into $n$ small intervals $[x_{i-1}, x_i]$ and the length of each interval is $\Delta x_i = x_i - x_{i-1}$. A Riemann sum for $f(x)$ is an expression

$$\sum_{i=1}^{n} f(\xi_i) \Delta x_i,$$

where $\xi_i$ are arbitrarily chosen numbers in $[x_{i-1}, x_i] (i=1,2,\cdots,n)$.

We say that $f(x)$ is Riemann-integrable on $[a,b]$ if there exists a real number $I$ with the following property: for $\forall \varepsilon > 0, \exists \delta(\varepsilon) > 0,$ such that

$$\left| \sum_{i=1}^{n} f(x) \Delta x_i - I \right| < \varepsilon$$

for all partitions $\sigma$ with $\sigma = \max_{1 \le i \le n} \{\Delta x_i\}_{n-1} < \delta$ and for any choice of $\xi_i \in [x_{i-1}, x_i]$. The above found number $I$ is the Riemann-integral of $f(x)$ over $[a,b]$, we denote

$$I = \int_a^b f(x) dx,$$

The idea of Riemann integral is making any partitions $\sigma$ in $[a,b]$ and constructing Riemann sum, then take limitation over $\sigma$ for the Riemann sum is the integral $\int_a^b f(x) dx$.

The Riemann integral permits a precise definition of the geometrical concept of "area" under a curve. So Riemann integral is an important tool in computing "area".

**2) Lebesgue integral**

The main difference between the Riemann and Lebesgue integrals is that the former uses intervals and their lengths while the latter uses more general point sets and their measures. Thus it is not surprising that Lebesgue integrals is more general than the Riemann integral.

First, we introduce the definition of Lebesgue measure. We denote $I$ as the open set

$$\{(x_1, x_2, \cdots, x_n) | a_i < x_i < b_i, \forall i=1,2,\cdots,n\}$$

in $R^n$ and $|I|$ as the volume of $I$. For any set $E \in R^n$, we define the measure of set $E$ to be the minimum of the sums of the volumes of families of open sets which cover $E$. To make this precise, we define the Lebesgue outer measure $m^* E$ as following

$$m^*E \stackrel{\Delta}{=} \inf\left\{\sum_i |I_i|, \{I_i\} \text{ is countable such that } E \subset \bigcup_i I_i\right\}$$

Similarly, we can define the Lebesgue inner measure $m_*E$ as

$$m_*E \stackrel{\Delta}{=} \sup\left\{\sum_i |I_i|, \{I_i\} \text{ is countable such that } \bigcup_i I_i \subset E\right\}$$

If $m^*E = m_*E$, we say $E$ is Lebesgue measurable.

Next, we introduce the definition of Lebesgue integral. Assume $E$ is a measurable set, $mE < +\infty$, $f(x)$ is a bounded function over $E$.

$$A \stackrel{\Delta}{=} \sup\{f(x), x \in E\}, \quad B \stackrel{\Delta}{=} \inf\{f(x), x \in E\}$$

Suppose a dissection of the interval $[A, B]$

$$A = y_0 < y_1 < y_2 < \cdots < y_{n-1} < y_n = B$$

Denote $E_i = \{x | y_{i-1} \leq f(x) \leq y_i\}$. For any $\xi_i \in [y_{i-1}, y_i] (i = 1, 2, \cdots, n)$, take summation

$$S(f, D) = \sum_{i=0}^{n} \xi_i mE_i$$

We say that $f(x)$ is Lebesgue-integrable on $E$ if $\lim_{\lambda \to 0} S(f, D)$ exits, where

$$\lambda = \max\{|y_i - y_{i-1}| | 1 \leq i \leq n\}.$$

We denote

$$\int_E f(x)dx = \lim_{\lambda \to 0} S(f, D)$$

We could describe Lebesgue integral using Fig. 8.

In order to obtain the area bounded by the curves $y = f(x)$, $y = 0$, $x = a$, $x = b$, we could use the rectangles from the dissection of axis $y$,

$$(b-a)(y_1 - y_0) + [(x_4 - x_1) + (b - x_5)](y_2 - y_1) +$$
$$[(x_3 - x_2) + (b - x_6)](y_3 - y_2) + (b - x_7)(y_4 - y_3).$$

When $\lambda = \max\{|y_i - y_{i-1}| | 1 \leq i \leq n\} \to 0$, then the summations of the areas of the rectangular is the Lebesgue integral of $f(x)$.

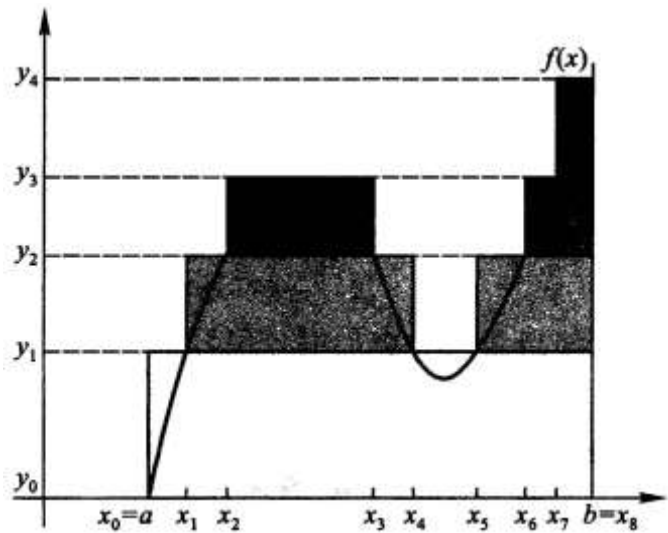

**Fig. 8**